# Time is Money: The Equilibrium Trading Horizon and Optimal Arrival Price

Kevin Darby, CQG Inc
kdarby@cqg.com

March 14, 2021

DRAFT COPY

**Abstract**

Executing even moderately large derivatives orders can be expensive and risky; it's hard to balance the uncertainty of working an order over time versus paying a liquidity premium for immediate execution.  Here, we introduce the Time Is Money model, which calculates the Equilibrium Trading Horizon over which to execute an order within the adversarial forces of variance risk and liquidity premium.  We construct a hypothetical at-the-money option within Arithmetic Brownian Motion and invert the Bachelier model to compute an inflection point between implied variance and liquidity cost as governed by a central limit order book, each in real time as they evolve.  As a result, we demonstrate a novel, continuous-time Arrival Price framework.  Further, we argue that traders should be indifferent to choosing between variance risk and liquidity cost, unless they have a predetermined bias or an exogenous position with a convex payoff.  We, therefore, introduce half-life factor asymptotics to the model based on a convexity factor and compare results to existing models. We also describe a specialization of the model for trading a basket of correlated instruments, as exemplified by a futures calendar spread.  Finally, we establish groundwork for microstructure optimizations within the Equilibrium Horizon framework.

## 1) The Trading Model

We consider the task of liquidating[1] a position $X_0$ optimally over a maximum time horizon T.

To be considered optimal, our solution must minimize deviation (or cost) from a benchmark; for the purposes of this study, the benchmark is $S_0$, the price of the asset at the instant we receive the order to liquidate.  This benchmark is often called Arrival Price.[2]

However, executing a sufficiently large order at $S_0$ is often impossible due to market mechanics and liquidity concerns, so we add an upper bound to our metric, which we call Sweep To Fill. Sweep To Fill is the effective price we would receive by executing the entire liquidation immediately as a market order (Perold 1988).

---

[1] An accumulation can also be considered in much the same manner.
[2] In practice, Arrival Price is defined as the midpoint between best bid and ask or some weighted formulation therein. (Wood 2011)

We construct the trading model under the auspices of Arithmetic Brownian Motion[3] (Bachelier 1900), in which the asset price S evolves via the stochastic differential equation:

$$dS_t = r \cdot S_t \, dt + \sigma dW_t$$

Where r is the risk-free interest rate, sigma ($\sigma$) is the annualized volatility in dollar terms and $W_t$ is the standard Brownian Motion (or Wiener process).

Given the SDE above, for 0 <= t <= T, the terminal value of S is:

$$S_T = S_t e^{r(T-t)} + \sigma \int_t^T e^{r(T-s)} \, dW_s$$

The distance between the initial and terminal values of S over the duration of the order is of primary concern to an execution trader. In an ideal (liquid, non-volatile) world, we could execute at a price exactly equal to $S_0$ (thereby mitigating any variance risk and paying no liquidity premium) at the instant we receive the signal to place the order.

In reality, however, we could perhaps get close to the Arrival Price benchmark for sufficiently small orders, but the liquidity structure of futures markets creates cost (slippage) and risk in the execution of larger orders because of the Central Limit Order Book.

## 2) The Central Limit Order Book (CLOB) and Liquidity Premium

Listed futures contracts are normally advertised to consumers in a structure resemblant of the figure below. Bid prices and sizes represent commitments to buy a particular amount of an instrument at a particular price; conversely, offers confer a commitment to sell. Each price level is an amalgam of constituent orders in agreement to transact at a single price.

This structure presents liquidity to the consumer, who either must be willing to pay a little more or sell for a little less to execute an aggressive order quickly, or wait in the queue to transact with an aggressive order that may or may not arrive later. For large orders, however, this immediate execution *premium* can be appreciable, yet executing the order over time is wrought with uncertainty.

**So the question is, how long should we work an order?**

---

[3] As we are primarily concerned with small trading horizons of a day or perhaps a week, ABM (Arithmetic Brownian Motion) is functionally equivalent to GBM, it's geometric and more famous counterpart (Almgren and Chriss 2000) (Gatheral and Schied 2011).

An Example of Liquidity Premium:

In the figure below, to liquidate 100 futures, the total cost of liquidity over $S_0$ is 44.5 [ 18*(0.125) + 36*(0.125+0.25) + 46*(0.125+0.25+0.25) = 44.5, or 0.445 per lot], in dollar terms this is an implicit cost of $2,225.00, or about 0.01% of the transaction. This effective immediate fill price (Sweep To Fill) can also be thought of as ArrivalPrice +/- LiquidityPremium, the cost of liquidity incurred to execute immediately.

$$LiquidityPremium = |SweepToFill - ArrivalPrice|$$

Together, this CLOB structure and the SDE above beget the dichotomy of liquidity premium and variance risk, for which we will now derive an equilibrium.

### 3) Options and the Half-Straddle

If, rather than paying the liquidity premium to execute immediately, we choose the path of variance risk, we can quantify that risk, the expected move[4] in S over the life of the order:

$$Exp[dS] = 2.0 \cdot \sigma\sqrt{(T-t)} / \sqrt{2\pi}$$

The equation above is also equal to the theoretical value of an at-the-money straddle with an expiration at time T, the maximum horizon of our order. This is to say that between $T_0$ and T, we would expect S to vary by this amount.

Now, we construct a hypothetical option (a half-straddle, struck at $K=S_0$) within Arithmetic Brownian Motion using the Bachelier option model (Grunspan 2011). This option's value will be half the expected move of S during the maximum horizon of our order (above); *it is the cost of insuring against a move in one direction.*

The price of a Vanilla Call Option[5] can be expressed as:

$$C_t = e^{-r(T-t)} ((S_T - k) \cdot \Phi(d_1) + g)$$

Where

$$d_1 = (S_T - k) / (\sigma\sqrt{(T-t)})$$

And

$$g = \varphi(d_1) \cdot \sigma\sqrt{(T-t)} \qquad (eq.\ 1)$$

Where $S_T$ is the forward price of the asset, r is the risk-free rate, k is the strike price, and $\Phi$ and $\varphi$ are the cumulative normal and probability density functions, respectively. Ignoring interest rates (over such a short period of time), we set r = 0, and $S_T = S_0$.

Then, with a little algebra, we find the value of the half-straddle (ATM Option with $S_T = k$) is:

$$H_t = e^{-r(T-t)} \varphi(0.0) \cdot \sigma\sqrt{(T-t)}$$

And recognizing

$$\varphi(0.0) = 1/\sqrt{2\pi}$$

We're left with:

$$H_t = \sigma\sqrt{(T-t)} / \sqrt{2\pi} = 0.5\ Exp[dS] \qquad (eq.\ 2)$$

---

[4] Net of interest rates
[5] Technically a liquidation or sell order would require a put, but since their prices are equal with ($S_T = k$) we defer to the longstanding practice of doing the math using the call.

### 4) The Martingale and The Equilibrium Trading Horizon

Given the dynamics described above, we now equate the (CLOB-based) liquidity premium to the premium of the hypothetical half-straddle[6] struck at Arrival Price ($S_0$). This is to say that the CLOB structure imbues the trader with *optionality* - a choice of either taking variance risk in executing the trade over *some amount of time*, or paying the liquidity *premium* instantaneously to mitigate that risk.

Now, if we know the instantaneous cost of liquidity, and if we can compute implied sigma from an estimate of spot volatility (Balta and Kosowski 2012)[7] and forward volatility through a liquid options market[8] (Jackel 2017) (LeFloc'h 2016), we can calculate the maximum tenor of a half-straddle purchasable with an investment of the liquidity premium-- this "implied time" is exactly equal to the future point in time at which variance risk equals the cost of liquidity.

Here, we compute the tenor of a half-straddle given a liquidity premium (L=H) and sigma:

$$H_t = L = (SweepToFill - ArrivalPrice) = LiquidityPremium$$

Where H(t) is the half-straddle (or ATM option):

$$L = H_t = \sigma\sqrt{(T-t)} / \sqrt{2\pi}$$

Given $\sigma$ we can solve for (T-t) and obtain:

$$T^* = (T-t) = 2\pi \cdot (H/\sigma)^2 \qquad \text{(eq. 3)}$$

Thus, if the fair value of the half-straddle is L and dollar volatility is $\sigma$, then the tenor of the half-straddle is T* above.

The martingale argument would suggest that if a trader would spend exactly this much time working the order, they would expose themselves to variance risk equivalent to the present value of paying the cost of the liquidity premium.

Thus **T\* is the Equilibrium Trading Horizon**, where variance risk and liquidity premium are equal, given an instantaneous cost (L=H) and volatility ($\sigma$).

---

[6] Given that an order is only ever a buy or a sell and not both, and that 50% of moves would be in our favor, we consider a half-straddle. Other structures (straddle, risk-reversal) could be considered based on risk tolerance and trading goals (Thanks to Scott Layne for this insight). See also Appendix 2.
[7] Also Parkinson (1980), Garman and Klass (1980), Yang and Zhang (2000), Rogers and Satchell (1991)
[8] Through an arbitrage free term-structure interpolation

### 5) Continuous-Time Arrival Price

Once we arrive at T*, we use it as an upper bound on the amount of time over which to execute our liquidation.

Based on this framework, a risk-neutral trader would execute an Arrival Price order (benchmarked between ArrivalPrice and SweepToFill), linearly[9] over duration [0, T*], the Equilibrium Horizon.

If the cost of liquidity and variance measures remain constant from $T_0$ to T* (and the trader is risk-neutral) then they can adhere to this initial projection as a static trading schedule. In holding T* constant, we arrive at a linear liquidation program where $X_0$ is the size of the position to liquidate and $X_t$ is the prescribed size of the position at time t:

$$T^* = 2\pi \cdot (L/\sigma)^2$$
$$X_t = X_0 \cdot (1 - t/T^*) \qquad \text{(eq. 4)}$$

However, variance and cost of liquidity will vary over the life of the order, consequently lengthening or shortening the equilibrium horizon as both measures evolve, the result of which gives rise to a continuous-time formulation as follows:

$$T_t^* = 2\pi \cdot (L_t/\sigma_t)^2$$
$$Z_t = X_0 \cdot (1 - t/T_t^*) \ ^{10}$$
$$X_t = Min[Z_t, X_{t-1}] \qquad \text{(eq. 5)}$$

Intuitively, this model exhibits adaptive qualities that are favorable in an Arrival Price program:
- As L is the cost of liquidity *in excess of* $S_0$, the model will increase execution speed as the market moves in our favor[11]
- If a large block of liquidity enters the market, the model will speed up to take liquidity
- If volatility decreases, such as after some event, the model will slow execution speed to wait for more liquidity to arrive

---

[9] Without consideration of microstructure, see section 9.
[10] An argument can be made for $Z_t = f(X_t)$ here, but the results are similar and the original form is simpler.
[11] Because liquidity would seem cheaper.

Here we see the continuous-time linear form in eq. 5 adapt to irregular liquidity.

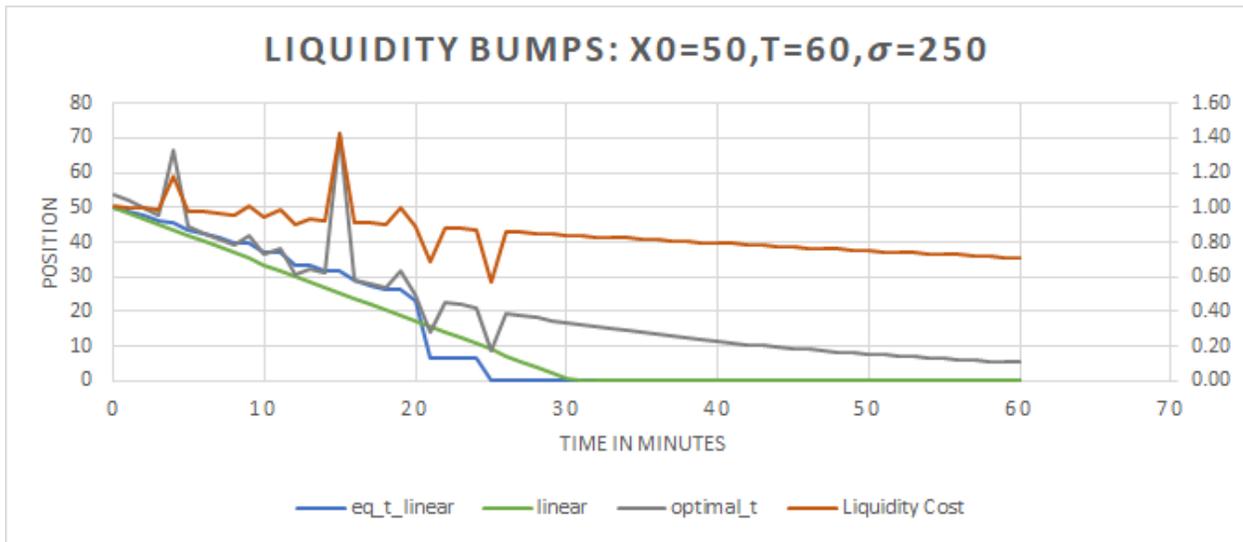

## 6) On Gamma and Bias

Above we assume that a trader has no preference between variance risk and liquidity cost. In practice, an execution trader may be inherently biased against volatility (Almgren and Chriss 2000) and would wish to execute a larger part of the order closer to the beginning of the Equilibrium Horizon.

Conversely, a trader may be trying to hedge a long gamma position, or have some other reason to prefer 'letting it ride' and do the bulk of his execution towards the end of the Horizon [12].

We introduce a parameter, *gamma* ($\gamma$), through the notion of a trader with an exogenous position that will experience a convex payoff causing them to either prefer or avoid price variance in execution.

Again, we visit the Bachelier model to compute first and second derivatives of Call Price:

$$\Delta = \frac{dC_t}{dS} = e^{-r(T-t)} \cdot \Phi(d_1)$$

$$\Gamma = \frac{d_2 C_t}{dS^2} = e^{-r(T-t)} \cdot \varphi(d_1) / (\sigma\sqrt{T-t})$$

Where

$$d_1 = (S_T - k) / (\sigma\sqrt{T-t})$$

And $\Phi$ and $\varphi$ are the cumulative normal and probability density functions, respectively.

---

[12] Pushing execution half-life toward the end of the horizon is sometimes called Destination Price

Bias Framework:

There are several ways to create a bias framework within the boundaries of the Equilibrium Horizon; here we explore a hyperbolic solution through which we can fix a half-life in [0,T*] and use units of ATM Option Gamma as a bias coefficient.

We consider now not only a liquidation of $X_0$, but also that the broader position contains some units of gamma[13]. First, we compute the expected gamma profit ($P_g$) over the duration of the order:

$$Exp[dS] = 2.0 \cdot \sigma\sqrt{(T-t)} / \sqrt{2\pi}$$
$$P_g = 0.5 Y \cdot \Gamma \cdot Exp[dS]^2$$
$$P_g = Y \ (\sigma\sqrt{(T-t)}) / (\sqrt{2} \cdot \pi^{3/2}) \qquad \text{(eq. 6)}$$

Where $Y$ is the quantity we created to mean units of ATM gamma.[14]

Below we look at a case of negative expected gamma profit and intuitively suggest that exogenous gamma costs equal to the cost of liquidity should accelerate trading within the Equilibrium Horizon, such that the half-life of the liquidation occurs approximately ¼ of the way through the initial outlook[15].

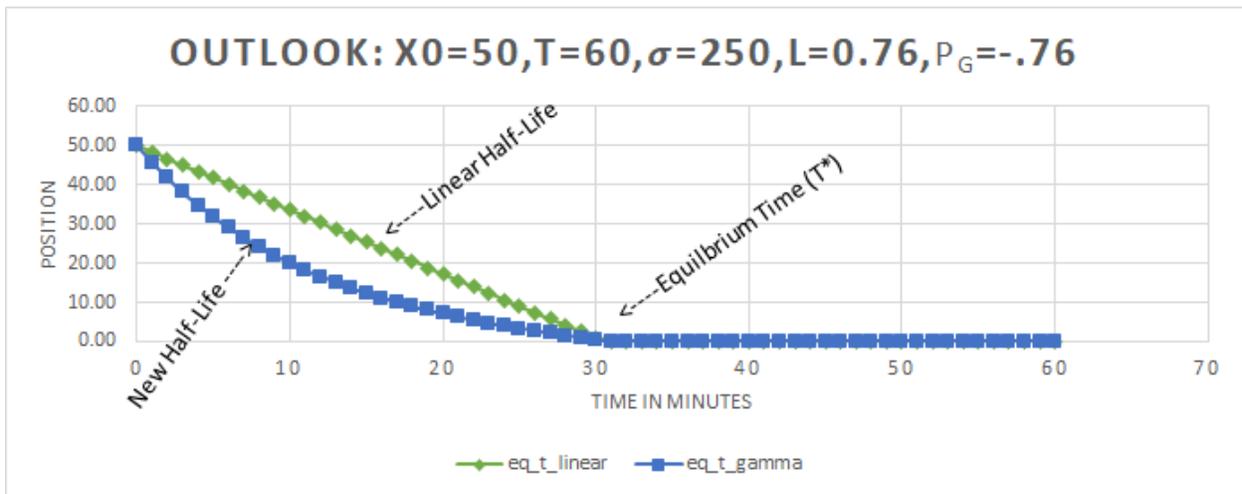

---

[13] Real or imagined
[14] Doing this in terms of ATM options allows for interesting scaling properties with sigma and t. It could also be defined in terms of raw $P_g$, or some other factor according to user preference.
[15] This ¼ number makes intuitive sense, but any method of determining half-life would suffice.

To integrate this bias framework into the Equilibrium Horizon, we reformulate equation 5 with hyperbolics:

$$A_g = P_{g(t)} / L_t$$
$$k_t = (1 + |A_g|)^{1/ln(2)} / T_t^*$$

For $A_g <= 0.0$

$$Z_t = Sinh(\ k_t\ (T_t^* - t)\ ) / Sinh(k \cdot T_t^*)$$
$$X_t = Min\ [\ X_0 Z_t\ ,\ X_{t-1}\ ] \qquad \text{(eq. 7)}$$

### 7) Calendar Spread Specialization

So far we've examined liquidating an outright futures position, and also a position with exogenous convexity. While a completely general framework for liquidating a correlated portfolio is reserved for future work, it's worthwhile to explore a specialization of the model for listed futures calendar spreads, the popular instruments used to roll commodity hedges in US Treasury and other complexes.

We consider equation 1, and redefine the Half-Straddle as a calendar spread option (Schaefer 2002):

$$C_{spd} = e^{-r(T-t)}\ (m\_t \cdot d_{spd} \cdot \Phi(d_{spd})\ +\ g)$$

Where

$$m = \sqrt{\sigma_1^2 + \sigma_2^2 - 2\dot{\rho}\sigma_1\sigma_2}$$
$$m\_t = m\sqrt{(T-t)}$$
$$d_{spd} = (S_{T1} - S_{T2} - k) / m\_t$$
$$g = \varphi(d_{spd}) \cdot m\_t \qquad \text{(eq. 8)}$$

Here $S_{T1}$ is the forward price of the first contract, $S_{T2}$ is the forward price of the second, and $\sigma_1$ $\sigma_2$ are the Bachelier sigmas. $\dot{\rho}$ is the correlation coefficient.

Again, we find the value of the half-straddle (ATM Option with $S_T = k$) is:

$$H_{spd} = e^{-r(T-t)} \varphi(0.0) \cdot m\_t$$

And (net of interest rates) we arrive at a familiar:

$$H_t = m\_t_t / \sqrt{2\pi}$$
$$T_{spd}^* = 2\pi (L_t/m_t)^2 \qquad \text{(eq. 9)}$$

The gamma of the spread is:

$$\Gamma_{spd} = \frac{d2C_{spd}}{dS^2} = e^{-r(T-t)} \cdot \varphi(d_{spd}) / m\_t$$

And it yields the expected gamma profit:

$$Exp[d_{spd}] = 2.0 \cdot m\_t / \sqrt{2\pi}$$
$$P_{g(spd)} = 0.5 \cdot \gamma \cdot \Gamma \cdot Exp[d_{spd}]^2$$
$$P_{g(spd)} = \gamma \cdot \Gamma \cdot m\_t^2 / \pi$$

And finally:

$$A_{spd} = P_{g(spd)} / L_{spd}$$
$$k_t = (1 + |A_{spd}|)^{1/ln(2)} / T_{spd}^*$$

For $A_{spd} <= 0.0$[16]

$$Z_{spd} = Sinh(k_t (T_{spd}^* - t)) / Sinh(k \cdot T_{spd}^*)$$
$$X_{spd(t)} = Min [X_0 Z_{spd}, X_{spd(t-1)}] \qquad \text{(eq. 10)}$$

---

[16] Derivation of a functional form for $A_{spd} > 0.0$ is left to the reader (hint: try tanh).

## 8) Comparison with Existing Models

While variance is usually a rather slow-moving measure, save discrete movements around market events, posted liquidity is often irregular, especially in an increasing volatility regime. Generally speaking, the model tends to wait out periods of high liquidity premium and later accelerate into periods of lower liquidity cost, continually optimizing with each tick.

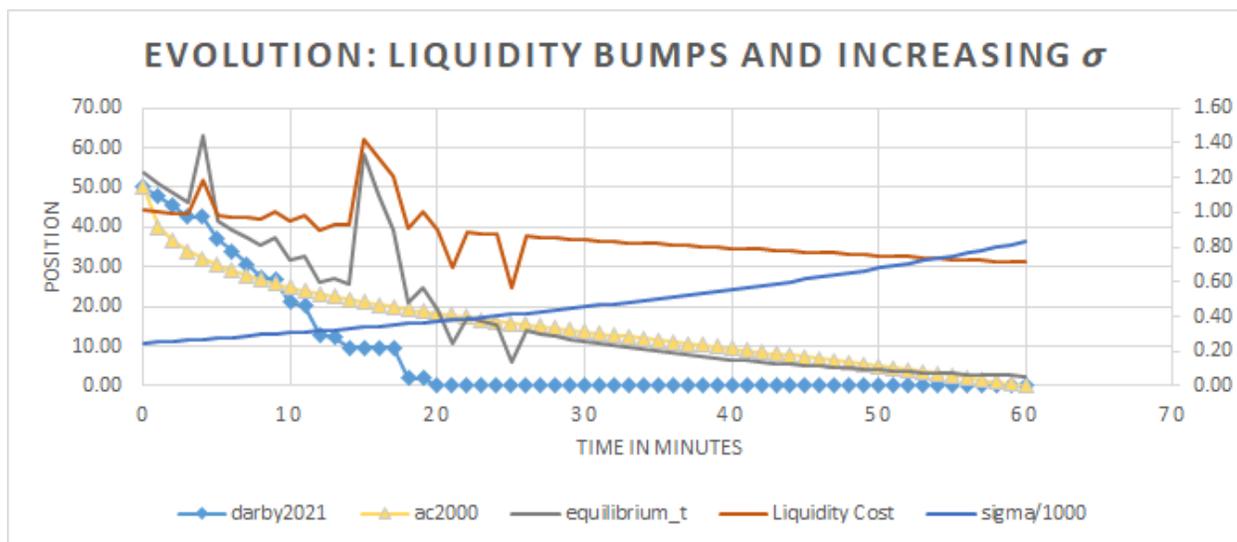

In the simulation above, we examine our model's performance as compared to a static AC2000 model (Almgren and Chriss 2000) during a 60-minute duration with liquidity cost spikes and rising volatility. As shown, equilibrium time varies widely over the life of the order, facilitating the effects shown. The models differ dramatically in that our model completes a full 40 minutes before static AC2000.

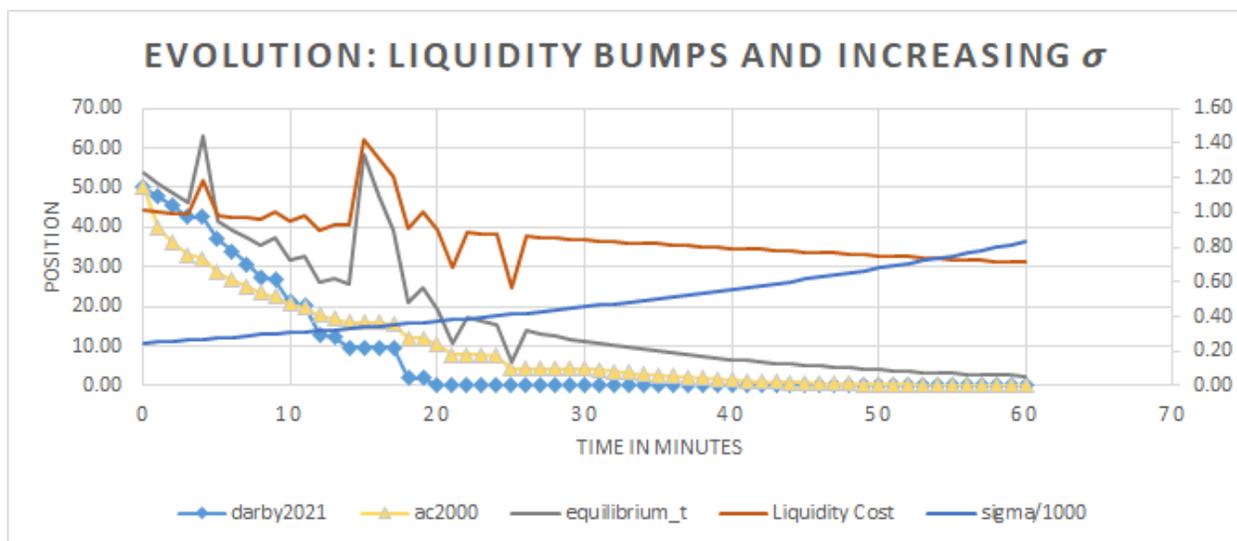

Above, we compare with a piecewise AC2000 model (Almgren and Lorenz 2007). The results are closer, but our model still completes 15 minutes before piecewise AC2000.

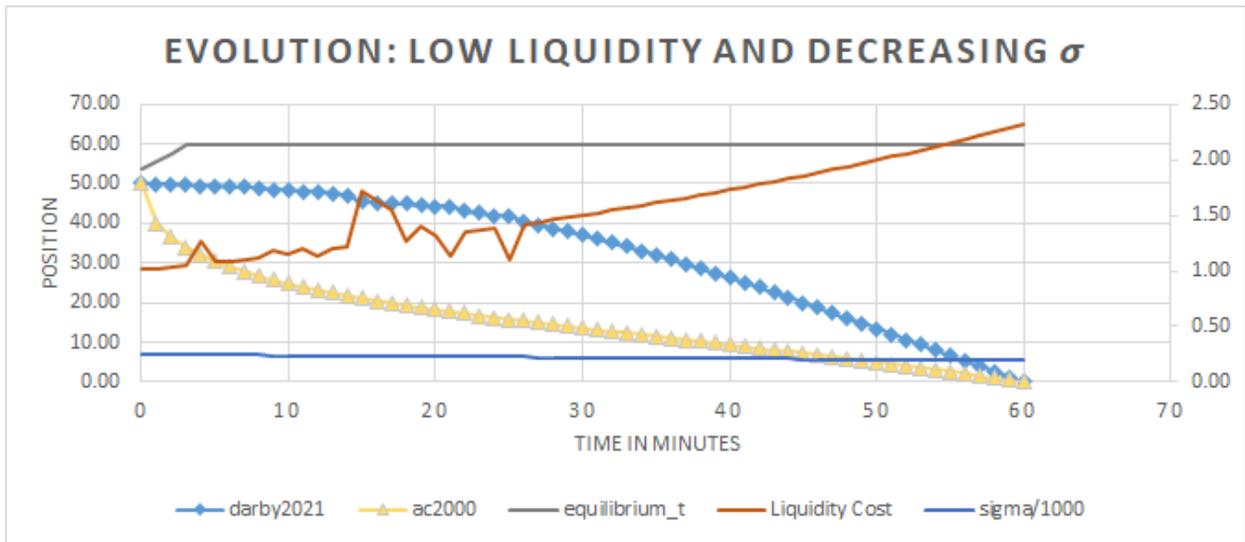

Here, we look at increasingly expensive, irregular liquidity and decreasing variance; the model is constrained by the maximum time of 60 minutes, but it waits as long as possible.

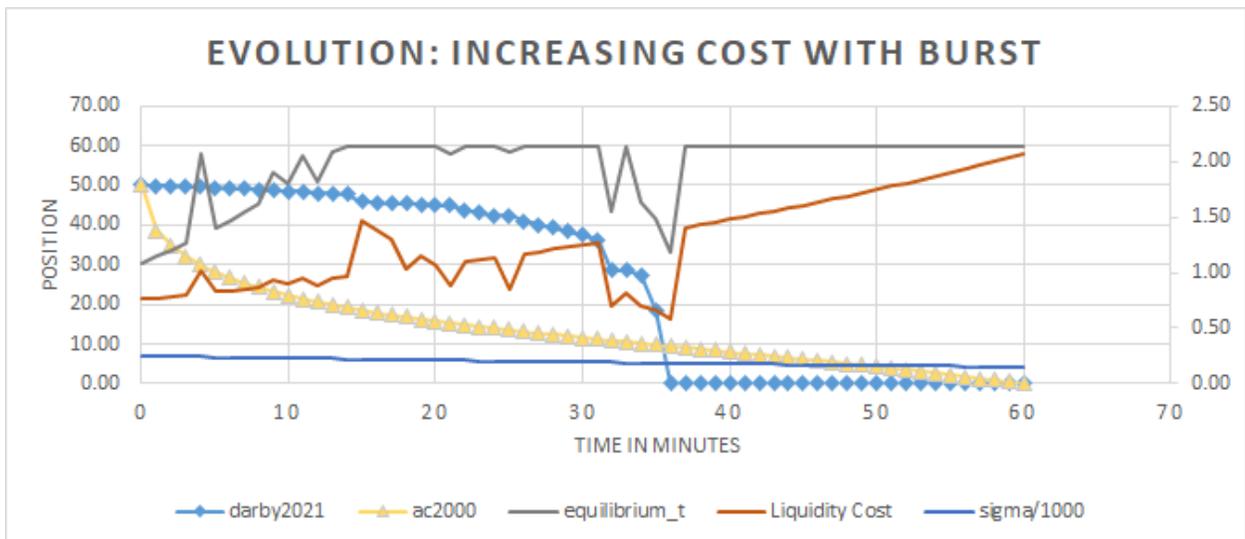

Finally, we demonstrate the model taking advantage of a sudden burst in liquidity to finish unexpectedly early.

### 9) But What About Microstructure?

The model is built around the macro-analytical precept of adversarial forces-- liquidity premium and variance risk, but the CLOB structure also creates opportunity and risk through the microstructure of the queues at each price level. While an in-depth analysis of market microstructure will be the subject of a future publication, it's worthwhile to examine an obvious case where microstructure dynamics modify the trajectory of the model in a manner consistent with the Equilibrium framework.

Below, we highlight two practical implementation details.

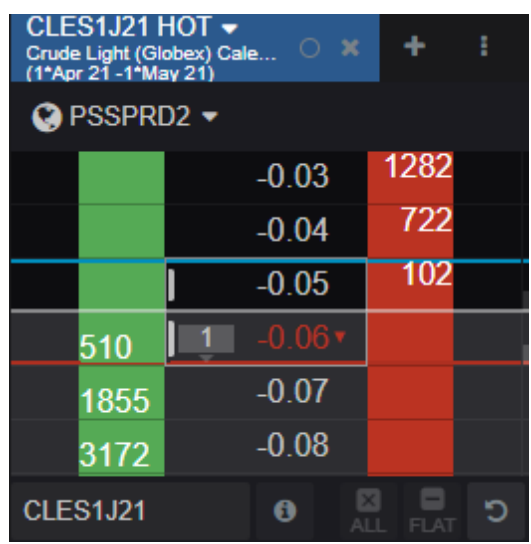

- Passive Liquidity

Providing passive liquidity when possible reduces the overall liquidity premium spent during execution. In the figure above, we see a large bid-side order imbalance that our algorithm would be wise to try to rectify. To this end, we optionally allow passive child orders to *extend* the Equilibrium Horizon, in as much as they correct an order imbalance or have a high probability of reducing overall liquidity cost. In our implementation, each passive child order contains logic to remain passive until microstructure no longer indicates a high probability of cost reduction.

- Optimal Size

It's also worthwhile to note that current market structure often prohibits placing large passive orders on the best bid or offer, as opposite liquidity often disappears as a result. Our algorithm uses a Market By Order Impact model to analyze order population metrics and compute optimal order size when sending a passive order. So, if we're trying to liquidate 100 futures spreads in a scenario similar to the above figure, we shouldn't just join the offer with a size of 100, unless an order of that size is sufficiently incognito. In this manner, the impact model may *restrict* the Equilibrium Horizon.

## 10) Backtest Results

To Be Published

## 11) Conclusion and Future Work

Here we introduced a novel, continuous-time Arrival Price framework based on the Equilibrium Trading Horizon which we derived from measures of variance risk and liquidity premium, as computed through market options prices and the limit order book, respectively.

We recognized that the limit order book structure imbues a trader with optionality and drew equivalence through options pricing theory, using listed options and spot-volatility measures as a proxy for short-term variance.

Further, we examined trader biases for or against the variance force and extended the linear model to a hyperbolic form to accommodate such biases; defining bias coefficient units as shares of ATM gamma. We also examined a specialization of the model for futures calendar spreads, honing in on a general form for a basket of correlated instruments.

Through comparison of this extended model to existing literature we found appreciable, beneficial differences to accepted dogma. **We answered the question, "How long should we work this order?"**

We look ahead to future work which will include an in-depth analysis of microstructure adaptations within the framework, through the lens of both standard methods and novel, machine-learning approaches. We will also examine alternative algorithmic implementations such as new options hedging methods and VWAP implementations, each grounded in this new risk-based Equilibrium Trading Horizon framework.

# Appendix

## A.1 On Short-Term Price Indicators and Drift

Until now we've ignored short-term interest rates; for rates in the realm of normal, and over reasonably short periods, this is probably a safe assumption. However, it's worthwhile to examine whether the model can be extended to accommodate short-term pricing trends in the form of 'drift' in the place of rates.

As such, we restate the central SDE as:

$$dS_t = \mu S_t \, dt + \sigma dW_t$$

Where μ is the coefficient that represents a short-term forward drift over the maximum tenor of the order. For the purposes of this discussion, we'll continue to ignore interest rates such that the present value of a dollar given at T is equal to one.

In keeping with our options framework, the expected change in S is still:

$$Exp[dS] = 2.0 \cdot \sigma\sqrt{(T-t)} / \sqrt{2\pi}$$

And the price of a Vanilla Put Option with F = $S_T$ is:

$$P_t = e^{-r(T-t)} \left( (k - F) \cdot \Phi(-d_1) + g \right)$$

Where

$$d_1 = (F - k) / (\sigma\sqrt{(T-t)})$$

And

$$g = \varphi(d_1) \cdot \sigma\sqrt{(T-t)} \qquad \text{(eq. 11)}$$

Where F is the forward 'drifted' value of the underlying price $S_T = F = e^{\mu(T-t)} S_0$

With F !=k, finding the "implied time" is more difficult, but we can conveniently use a slight modification of the Jäckel method of approximating implied normal volatility.

We define:

$$\tilde{\Phi}(x) := \Phi(x) + \varphi(x)/x$$

And

$$x := (k - F) / (\sigma\sqrt{T - t})$$

And Solve

$$\tilde{\Phi}(x) = P_t / (k - F)$$

Where $P_t$ is the value of the (OTM) put option above, thus:

$$T^* = (T - t) = [(k - F) / (x \cdot \sigma)]^2 \qquad \text{(eq. 12)}$$

Where x is found via the inversion described in (Jäckel 2017). Thus, the Equilibrium Trading Horizon is equal to the "implied time" of a hypothetical vanilla put option with $P_t$ = L for a liquidation, and in a similar fashion, the call for an accumulation trade.

## A.2 Conditional Expected Slippage

Above we modify the "half-straddle" notion of the model and define $S_T$ (or F) as a function of short-term drift and use Jäckel's inversion to calculate the Equilibrium Horizon as the "implied time" of an option with a value equal to the liquidity premium, L. In a similar fashion, we may redefine k such that the model reflects a maximum conditional expected slippage in the Equilibrium Horizon. Here we examine an accumulation trade and remember that the ensuing hypothetical call option is a *contingent claim.*

The call's payoff at maturity is the positive difference between the forward price F and the strike k because we receive the value of F in exchange for paying the strike price k at maturity:

$$C_T = Max(F - k, 0) = E[(F - k)_+]$$

Ignoring interest rates, we can decompose this for time t < T and write:

$$C_t = a + b$$

Where α is a credit of the expected value of F, given F > k, times the probability of {F>k}:

$$a = P\{F > k\} \cdot E[F \mid F > k]$$

And b is a debit of k times the probability of {F>k}:

$$b = P\{F > k\} \cdot -k$$

Using Bachelier's fundamental principle, as in Eq. 1:

$$P\{F > k\} = \Phi[(F - k)/(\sigma\sqrt{(T - t)})] = \Phi(d1) \quad \text{(eq. 13)}$$

The conditional expected value of F, given F > k is:

$$E[F + x \mid F > k] = F + \sigma\sqrt{(T - t)} \cdot \varphi(d1) = F + g \quad \text{(eq. 14)}$$

Which yields a form similar to eq. 1

$$C_t = F\,\Phi(d_1) + g - k\,\Phi(d_1)$$

Therefore, we can redefine our hypothetical half-straddle as an out-of-the-money option and make a new replication argument that allows for a quantifiable upper bound on slippage-- instead of k = $S_0$ = Arrival Price, we use k = Arrival Price + Slippage.

Thus "spending" liquidity premium L to hedge the conditional expectation of the contract rising above k, in the eventuality that F > k, yields an intuitively larger T* than the half-straddle case which can be computed reliably with the inversion described in Appendix 1.